\documentclass[prl,twocolumn,superscriptaddress,nopacs,amsmath,amssymb,nofootinbib]{revtex4-2}
\usepackage{mathptmx}
\usepackage[T1]{fontenc}
\usepackage[latin9]{inputenc}
\usepackage{float}
\usepackage{amssymb}
\usepackage{graphicx} 
\usepackage{esint}
\usepackage{hyperref}
\usepackage{color}
\usepackage{booktabs}
\usepackage{makecell}
\usepackage{soul}
\usepackage{lineno}

\makeatletter
\renewcommand\[{\begin{equation}}
\renewcommand\]{\end{equation}} 

\makeatother

\usepackage{babel}
\begin{document}
\title{Quantum critical phase of FeO spans conditions of Earth's lower mantle}
\author{Wai-Ga D. Ho}
\affiliation{Department of Physics and National High Magnetic Field Laboratory, Florida State University, Tallahassee, FL, USA}
\author{Peng Zhang}
\affiliation{MOE Key Laboratory for Non-equilibrium Synthesis and Modulation of Condensed Matter, Shaanxi Province Key Laboratory of Advanced Functional Materials and Mesoscopic Physics, School of Physics, Xi'an Jiaotong University, 710049, Xi'an, Shaanxi, P.R.China}
\author{Kristjan Haule}
\affiliation{Center for Materials Theory, Department of Physics, Rutgers University, Piscataway, NJ, USA}
\author{Jennifer M. Jackson}
\affiliation{Seismological Laboratory, California Institute of Technology, Pasadena, CA, USA}
\author{Vladimir Dobrosavljevi\'{c}}
\affiliation{Department of Physics and National High Magnetic Field Laboratory, Florida State University, Tallahassee, FL, USA}
\author{Vasilije V. Dobrosavljevic}
\affiliation{Seismological Laboratory, California Institute of Technology, Pasadena, CA, USA}
\affiliation{Earth and Planets Laboratory, Carnegie Institution for Science, Washington, D.C. 20015, USA}

\maketitle
{\bf Earth's interior consists primarily of an insulating rocky mantle \cite{Sinmyo2014,Ohta2017} and a metallic iron-dominant core \cite{Ohta2016,Pozzo2013}. Recent work has shown that mountain-scale structures at the core-mantle boundary may be highly enriched in FeO \cite{Wicks2010,Bower2011,Dobrosavljevic2019,Lai2022}, reported to exhibit high conductivity and metallic behavior at extreme pressure-temperature ($P$--$T$) conditions \cite{ohta2012}. However, the underlying electronic processes in FeO remain poorly understood and controversial. Here we systematically explore the electronic structure of $B$1-FeO at extreme conditions with large-scale theoretical modeling using state-of-the-art embedded dynamical mean field theory (eDMFT) \cite{haule2010}. Fine sampling of the phase diagram at more than 350 volume-temperature conditions reveals that, instead of sharp metallization, compression of FeO at high temperatures induces a gradual orbitally selective insulator-metal transition. Specifically, at $P$--$T$ conditions of the lower mantle, FeO exists in an intermediate "quantum critical" state, characteristic of strongly correlated electronic matter \cite{Terletska,Vucicevic2013,Furukawa2015}. Transport in this regime, distinct from insulating or metallic behavior, is marked by incoherent diffusion of electrons in the conducting $t_{2g}$ orbital and a band gap in the $e_g$ orbital, resulting in moderate electrical conductivity ($\sim 10^5$ S/m) with modest $P$--$T$ dependence as observed in experiments \cite{ohta2012}. FeO-rich regions in Earth's lowermost mantle could thus influence electromagnetic interactions between the mantle and the core, producing several features observed in Earth's rotation and magnetic field evolution \cite{Buffett2015}. }\vspace{12pt} 

{\em Introduction --}
Earth's lower mantle is thought to be composed primarily of bridgmanite (Mg$_{1-x}$Fe$_x$)SiO$_3$ and ferropericlase (Mg$_{1-x}$Fe$_x$)O, where $x \sim 0.1 - 0.2$, coexisting with CaSiO$_3$ \cite{Mattern2005,Irifune1994,Allegre1995}. These major mineral phases behave as insulating materials up to conditions of the lowermost mantle, with electrical conductivities on the order of $10^0$ to $10^2$ S/m \cite{Sinmyo2014,Ohta2017}, many orders of magnitude lower than proposed conductivities of the metallic iron-dominant core ($\sim 10^6$ S/m) (e.g., \cite{Ohta2016,Pozzo2013}). Instead of a homogeneous lower mantle, seismic observations over the last several decades have robustly identified multi-scale structures across Earth's core-mantle boundary \cite{Jackson2021,Sun2019}. These structures have been grouped into two main categories: (1) two continent-scale "large low-seismic velocity provinces" (LLSVPs), considered to be piles of heterogeneous material or bundles of thermochemically distinct mantle plumes \cite{Garnero2016,Tan2005}, and (2) numerous mountain-scale "ultralow velocity zones", basal structures discovered within and around the edges of LLSVPs, including at the roots of major mantle plumes like those that source volcanism at Hawai'i, Iceland, and the G\'alapagos \cite{Garnero1998,Wen1998,Yu2018,Kim2020,Yuan2017,Jenkins2021,Thorne2021,Cottaar2022}.

\begin{figure}[b]
\vspace{-12pt}
\includegraphics[width=3.5in]{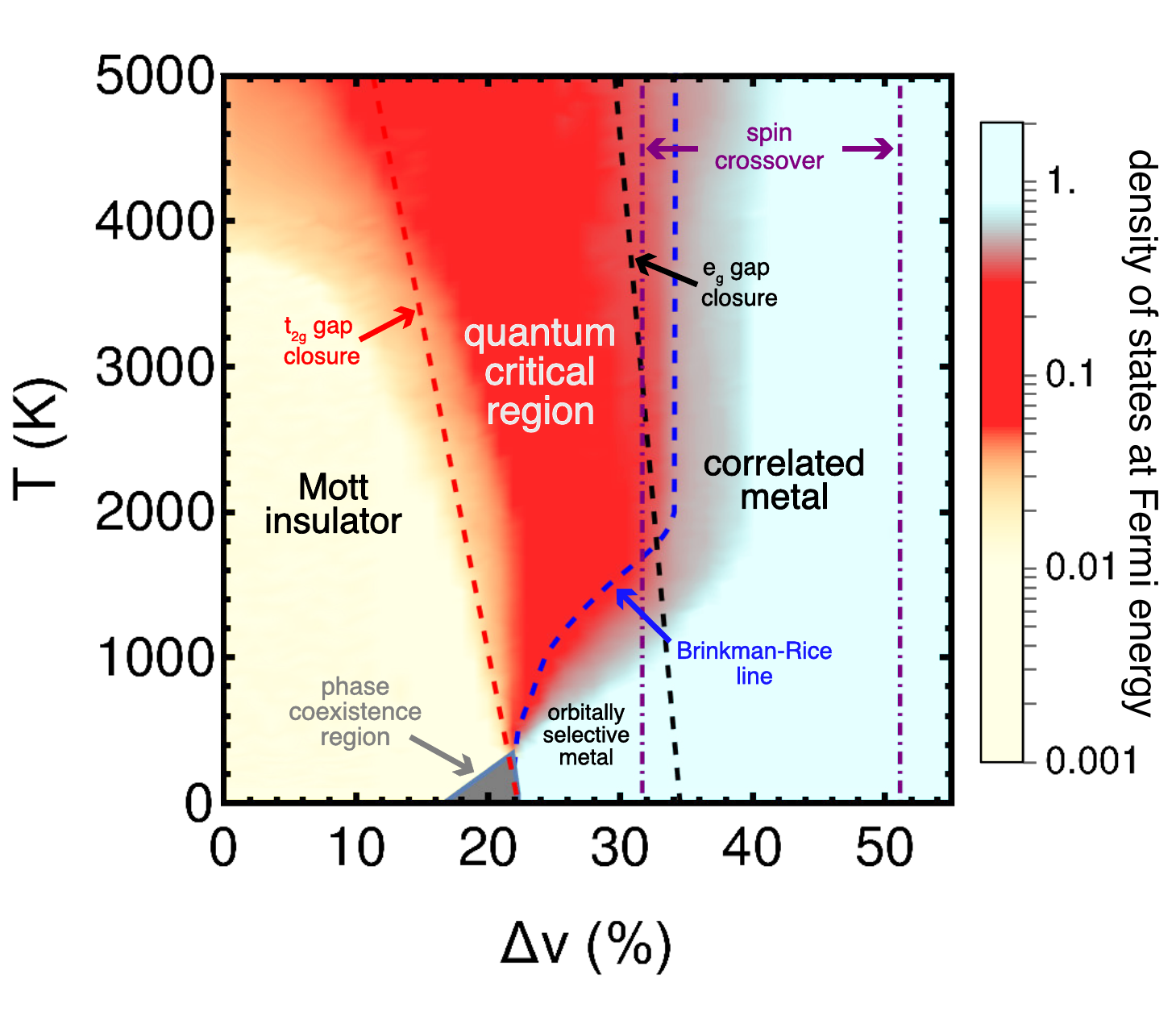}
 \caption{Theoretical phase diagram for the $B$1 (rocksalt) phase of FeO, as a function of the reduced volume $\Delta \rm{v} = (\rm{v}_o - \rm{v})/\rm{v}_o$ and temperature $T$, where $\rm{v}_o$ is the volume of FeO at ambient conditions. 
 The color-coded value of the electronic density of states (DOS) at the Fermi energy is used to distinguish the (gapped) Mott insulator from the metal and the intermediate "quantum critical" regime \cite{Terletska,Furukawa2015}. The "Brinkman-Rice" crossover line \cite{rqp2013prl} marks the thermal destruction of coherent quasiparticles in the metal with increasing temperature \cite{pustogow2021npjQM}. 
}
\label{DOSphasediagram}
\vspace{-12pt}
\end{figure}

Studies generally agree that the interpretation of these observed structures requires strong compositional contrasts from the surrounding average lower mantle and possibly the presence of partial melt \cite{Yu2018,McNamara2019,Williams1996}. Recent interdisciplinary work on ultralow velocity zones has demonstrated that solid FeO-rich mineral assemblages, consisting of iron-rich (Mg$_{1-x}$Fe$_x$)O  ($x \sim 0.8 - 0.95$) coexisting with (Mg,Fe)SiO$_3$ and CaSiO$_3$, can produce structures that satisfy the velocity reductions and topographies constrained by seismic observations and geodynamic simulations \cite{Wicks2010,Bower2011,Wicks2017,Dobrosavljevic2019,Lai2022}. Such strong iron enrichment, arising from crystallization of the primordial magma ocean or chemical interactions with the iron core, leads to several unique physical properties observed for the very iron-rich (Mg,Fe)O phase, including high seismic anisotropy \cite{Finkelstein2018}, remarkably low viscosity \cite{Reali2019}, and high electrical conductivity ($10^5$ to $10^6$ S/m) approaching those of a metallic material \cite{ohta2012,Ohta2014}. 

\begin{figure*}[t]
\includegraphics[width=7in]{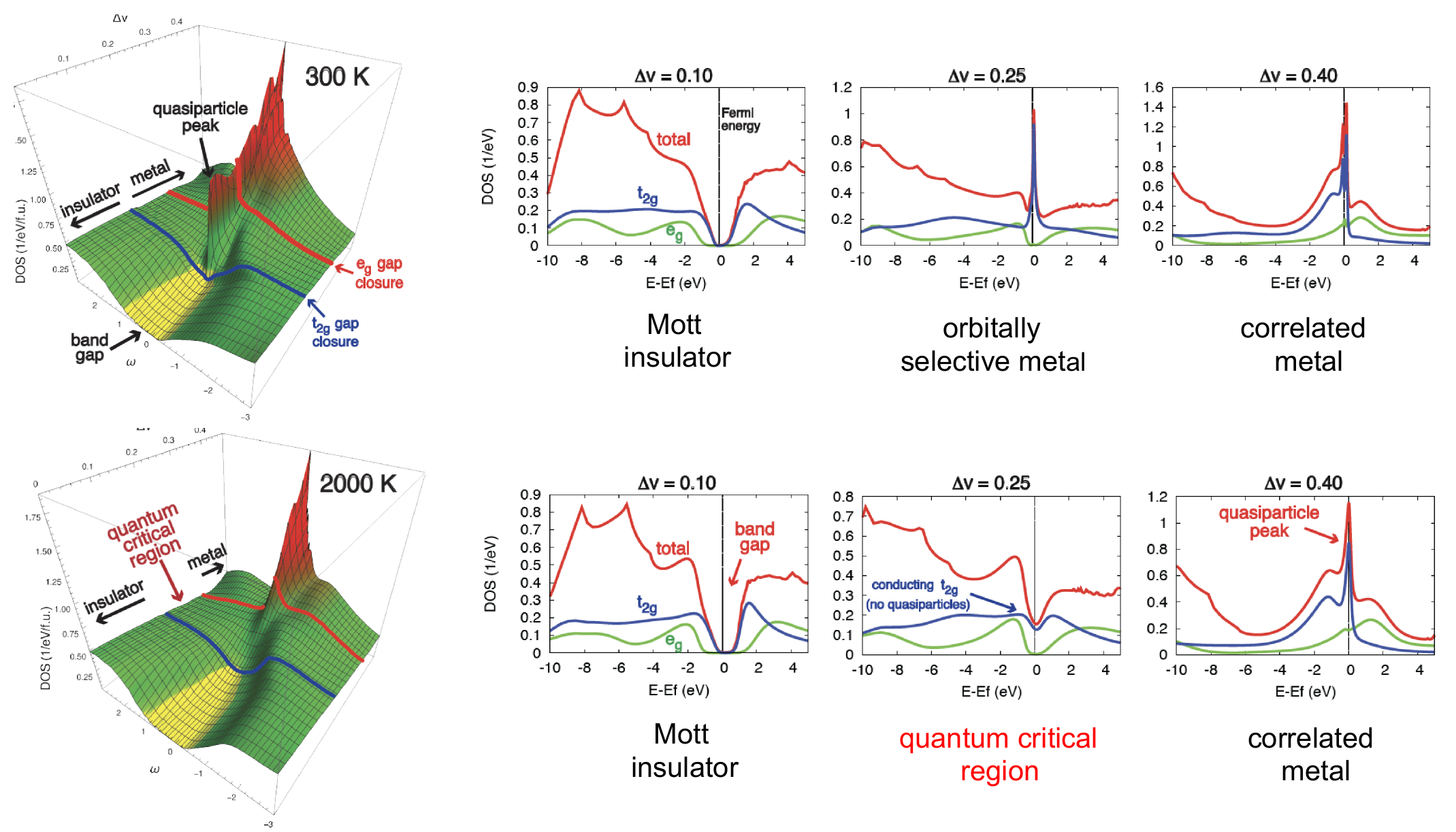}
\vspace{12pt}
\caption{Evolution of the electronic density of states (DOS) with compression. Metallization is sharp at low temperature ($T=300$ K, top row), where closure of the Mott gap in the $t_{2g}$ band leads to immediate emergence of coherent quasiparticle states at the Fermi energy. In contrast, a broad intermediate "quantum critical" phase arises at higher temperatures ($T$ = 2000 K, bottom row), with a spectral pseudogap (reduced but finite DOS) and no quasiparticle states, characteristic of an incoherent conductor. Here, the quasiparticle peak appears only at further compression with the closure of the $e_{g}$ gap and onset of the spin crossover phenomenon.
}
 \label{Vscan2T}
\end{figure*}

However, the electronic properties and related transitions in FeO and iron-rich (Mg,Fe)O remain poorly
understood and controversial. An insulator-metal transition has been proposed for FeO from measurements of relatively high conductivity ($\sim 10^5$ S/m) with weak $P$--$T$ dependence above $\sim$ 60 GPa \cite{ohta2012}, while similarly high conductivity was reported for (Mg$_{0.2}$Fe$_{0.8}$)O and (Mg$_{0.05}$Fe$_{0.95}$)O  but interpreted as insulating behavior up to $\sim$130 GPa \cite{Ohta2014}. 
Standard electronic-structure theory methods focus at $T=0$ K, and are not able to properly capture thermal effects, which often dominate in the vicinity of the insulator-metal transition \cite{mott1990metal,dobrosavljevic2012conductor}. Such extreme fragility of electronic states is especially pronounced in "strongly correlated" \cite{vollhardt2019jpspc} electronic systems \cite{Furukawa2015,dressel2020advphys,moire2021nature}, often featuring tightly-bound $d$ or $f$ orbitals \cite{tokura2000science}. Here the Coulomb repulsion between pairs of electrons confined to the same orbital takes center stage, typically resulting in very strong electron-electron scattering and poor conduction at elevated temperature \cite{MIR_limit}. Given these complications, several fundamental open questions arise regarding the insulator-metal transition (IMT) in $B$1-FeO at high pressures: (1) Is there a sharp IMT at high temperature, in the regime characteristic of Earth's deep mantle? (2) What is the mechanism of electronic transport (i.e., the dominant form of scattering) in this regime? (3) How do orbital selectivity \cite{vojta2010orbital,vojta2010orbital} and the associated spin-crossover affect the transition region? 

Knowledge of electronic processes in FeO at extreme conditions and consequences for transport
properties is essential for understanding phenomena at Earth's core-mantle boundary, including electromagnetic
coupling of the core and mantle and heat flow through this region. To that end, we employ a state-of-the-art "embedded DMFT" (eDMFT) ab initio approach \cite{haule2010} that combines dynamical mean field theory (DMFT) methods \cite{Georges1996,dmft2006rmp} and standard density functional theory (DFT) with full charge self-consistency. While some valuable steps in this direction have been taken in previous work \cite{shorikov2010prb,ohta2012,leonov2017prb,leonov2020prb}, sufficiently detailed and systematic study of the transition region has not been performed, preventing a clear understanding of the important open questions at hand. Using this approach, we systematically survey the electronic structure of cubic $B$1-FeO, the crystal structure relevant to Earth's lower mantle conditions \cite{fischer2011}. An expansive data set featuring calculations at more than 350 temperature-volume conditions (see Supplementary Materials) finely samples the phase diagram up to conditions of Earth's inner core (300 GPa, 5000 K). This detailed information allows us to accurately determine and physically interpret the boundaries of different transport regimes across the phase diagram. 

\vspace{12pt}

\noindent {\em Results}\vspace{12pt}

{\em Three distinct electronic phases of $B$1-FeO --} Our theoretical calculations reveal three distinct electronic phases in the high-$P$--$T$ phase diagram of $B$1-FeO (Fig.~\ref{DOSphasediagram}). At ambient conditions and low degrees of compression, FeO behaves as a Mott insulator, in which both the $t_{2g}$ and $e_g$ orbitals exhibit large band gaps at the Fermi energy on the order of several eV and electrons remain bound to their respective nuclei. In contrast, at large degrees of compression, FeO exists as a strongly correlated metal, where one or both the $d$ orbital band gaps are closed, producing a characteristic "quasiparticle" density of states (DOS) peak at the Fermi energy (see also Fig.~\ref{Vscan2T}, rightmost panels). These coherent quasiparticle states behave akin to those of weakly interacting electrons, allowing the low-energy charge excitation in a correlated metal to travel as coherent waves with low levels of scattering. 

\begin{figure}[t]
\includegraphics[width=2.8in]{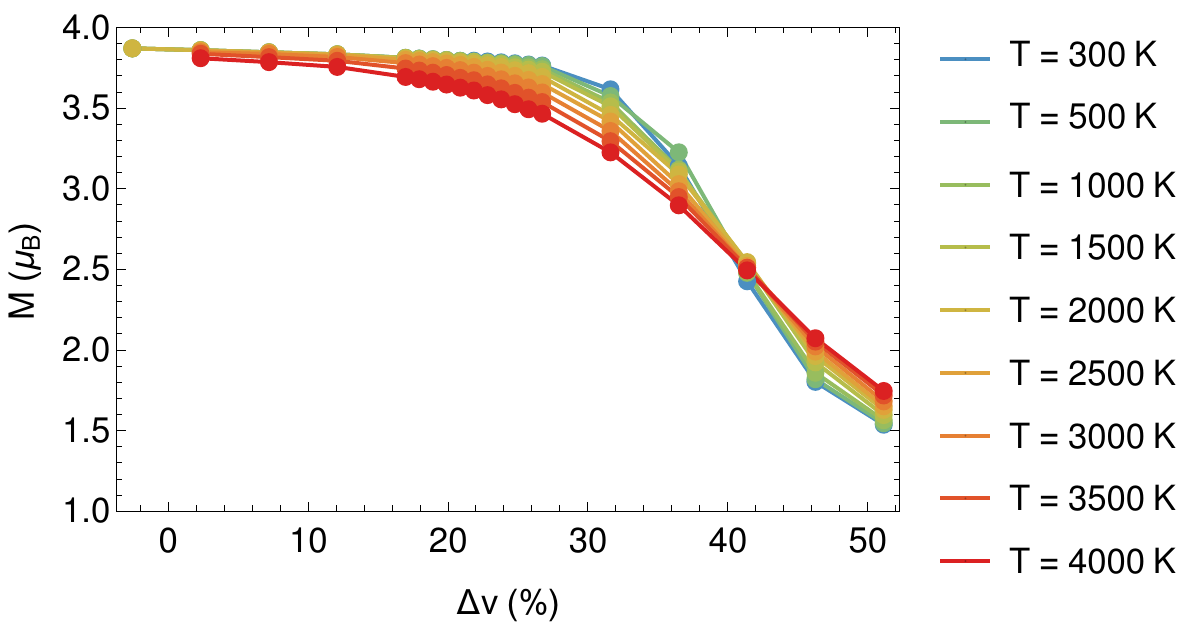}\\
\includegraphics[width=2.8in]{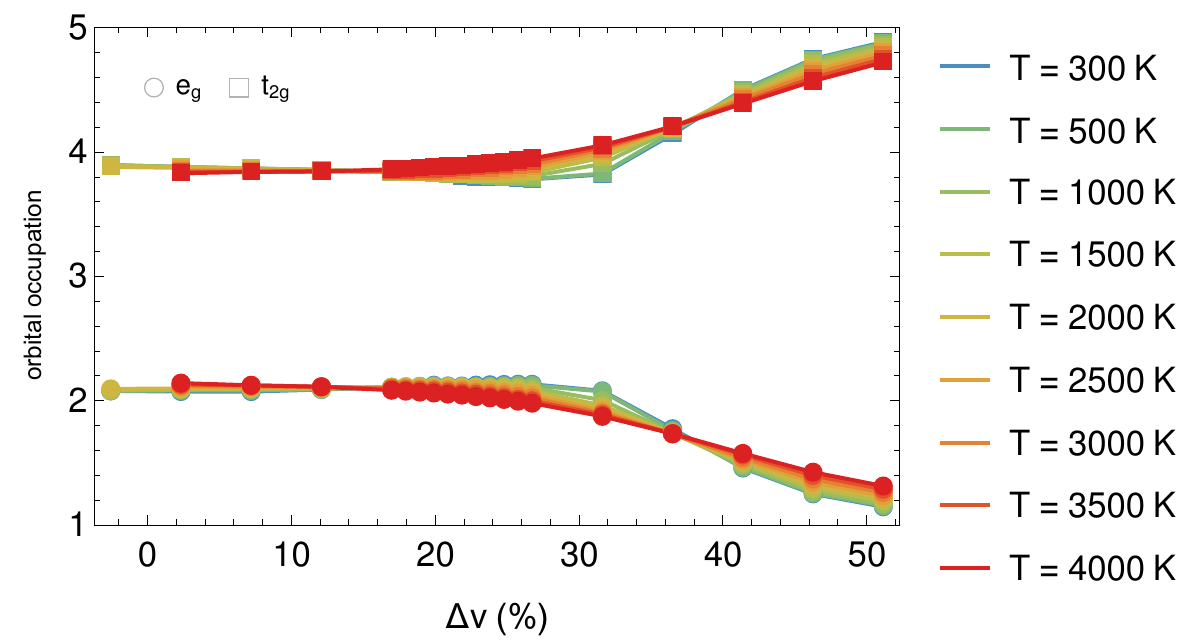}
\caption{Spin crossover behavior of FeO at high $P$--$T$ conditions. Magnetic moment $M$ results (top panel) show that  Fe remains in the high spin state throughout the insulator-metal transition region, with the onset of spin crossover only upon further compression with the closure of the $e_g$ gap. The spin crossover reflects partial charge transfer from the $e_g$ to the $t_{2g}$ orbital (bottom panel) and displays remarkably weak $T$-dependence. 
}
\label{crossover}
\end{figure}

At intermediate degrees of compression and sufficiently high temperatures, FeO exists in  a "quantum critical" (QC) state, which is notably different from either an insulator or a metal. Here, the $t_{2g}$ gap has closed to form a conducting band, but unlike in a conventional metal, the density of states at the Fermi energy is significantly reduced, with a marked absence of quasiparticles (Fig.~\ref{Vscan2T}, bottom row). Instead of traveling as coherent waves with minimal scattering as in a metal, electrons in the QC state exhibit incoherent diffusion marked by strong electron-electron scattering with a short mean-free path at the scale of atomic spacing. In this regime, the $e_g$ gap remains open and FeO remains in the high-spin state, with four $d$ electrons in the $t_{2g}$ orbital (Fig. \ref{crossover}). We stress that the QC phase arises only at finite temperatures above the insulator-metal phase coexistence region, terminating at the critical end-point $T_c \sim 370$ K; the insulator-metal transition assumes first-order character at $T < T_c$. \vspace{12pt}

{\em Temperature-dependent forms of the insulator-metal transition --} The physical nature of the insulator-metal transition in FeO and the range of pressures spanning the QC region depend strongly on the range of temperatures considered. 
At low temperatures ($T \leq T_c$),  FeO transitions directly from a Mott insulator to an "orbitally selective"  metal around $\Delta v \sim 20\%$ (corresponding to $P  \sim 58$ GPa \cite{fischer2011}). Here the closure of the $t_{2g}$ gap leads to the immediate formation of a quasiparticle peak at the Fermi energy in the $t_{2g}$ orbital (see Fig.~\ref{Vscan2T}, top row), while the $e_g$ gap remains open. These quasiparticle states are remarkably fragile to thermal excitations, and are suppressed around the "Brinkman-Rice" temperature $T_{BR}$ (see Fig.~\ref{DOSphasediagram}), marking the crossover to the QC phase. As $T_{BR}$ increases with compression, the insulator-metal transition is "smeared out", producing an increasingly wider QC  "fan" at $T_c < T \lesssim 2000$ K. The left boundary of the QC region corresponds to a temperature scale where the Mott gap is smeared through thermally activated processes (see Supplementary Materials for precise definition of the corresponding crossover lines shown in Fig.~\ref{DOSphasediagram}). 

This behavior becomes qualitatively different at very high temperatures. At $T \gtrsim 2000$ K, the quasiparticles are unable to form in the $t_{2g}$ orbital before compression causes the closure of the $e_g$ gap, around $\Delta v \sim 34\%$. Further compression leads to the onset of spin crossover phenomena and simultaneous formation of a  correlated metal, with robust quasiparticles forming in both sectors. The spin crossover extends over a wide compression range with weak temperature dependence (Fig.~\ref{crossover}), and is marked by a partial charge transfer from the $e_g$ to the $t_{2g}$ orbital, with one electron remaining in the $e_g$ orbital and a drop in the magnetic moment from 4 to $\sim 1.5$ Bohr magneton. Unlike $T \lesssim 2000$ K, where the QC region gradually broadens with increasing temperature, here the transition to a quasiparticle metal occurs immediately after the $e_g$ gap closure and spin crossover onset, leading to a Brinkman-Rice line with weak temperature dependence and an abridged pressure extent for the QC "fan" at high temperatures. Orbital selectivity and the associated spin crossover phenomena thus dramatically affect the form of the insulator-metal transition behavior at these very high temperatures, producing markedly weak temperature dependence of all physical quantities within the QC region. 

\begin{figure}[t]
\vspace{-12pt}
\begin{center}
\includegraphics[width=3.8in]{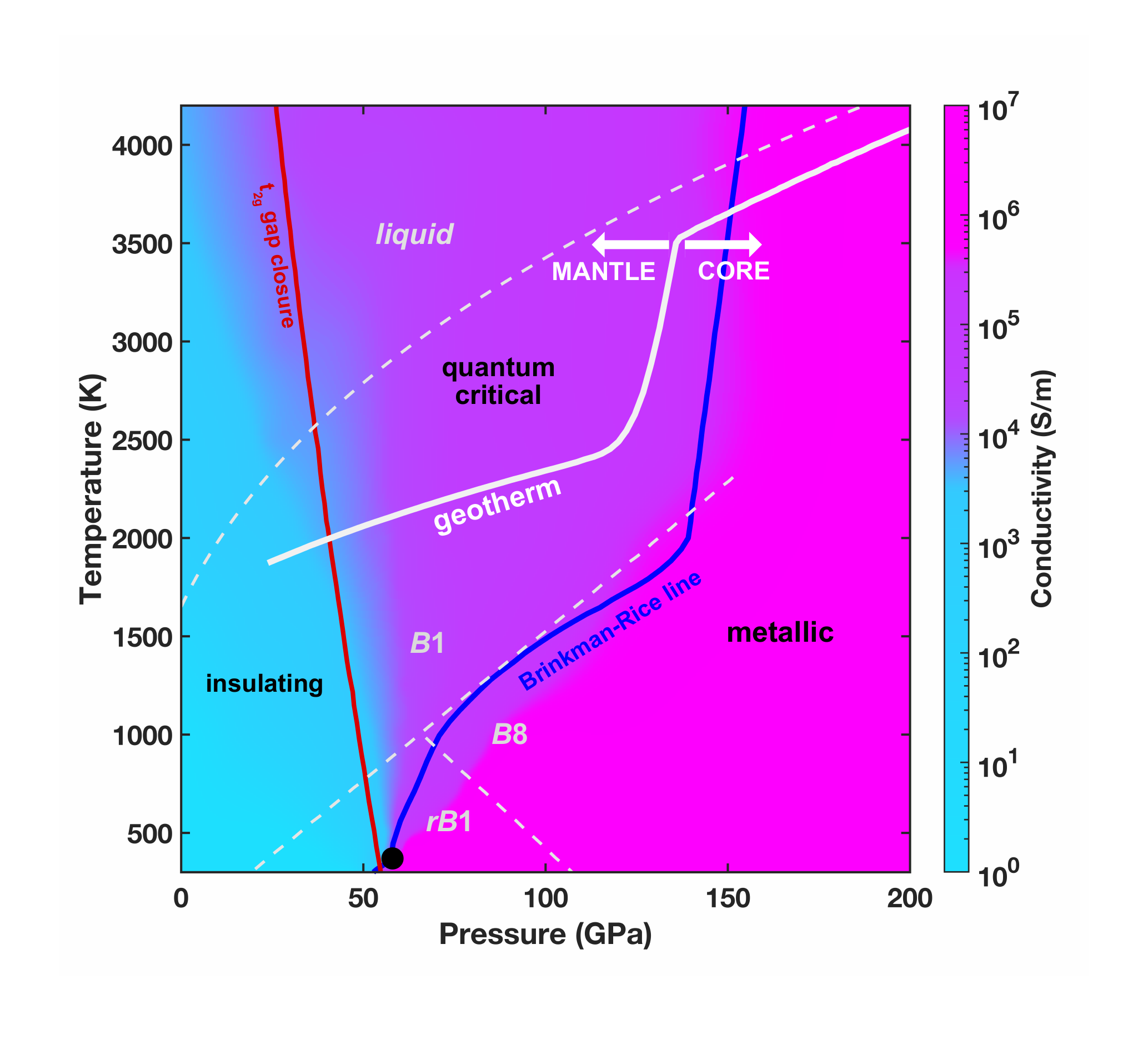}\\
\includegraphics[width=3in]{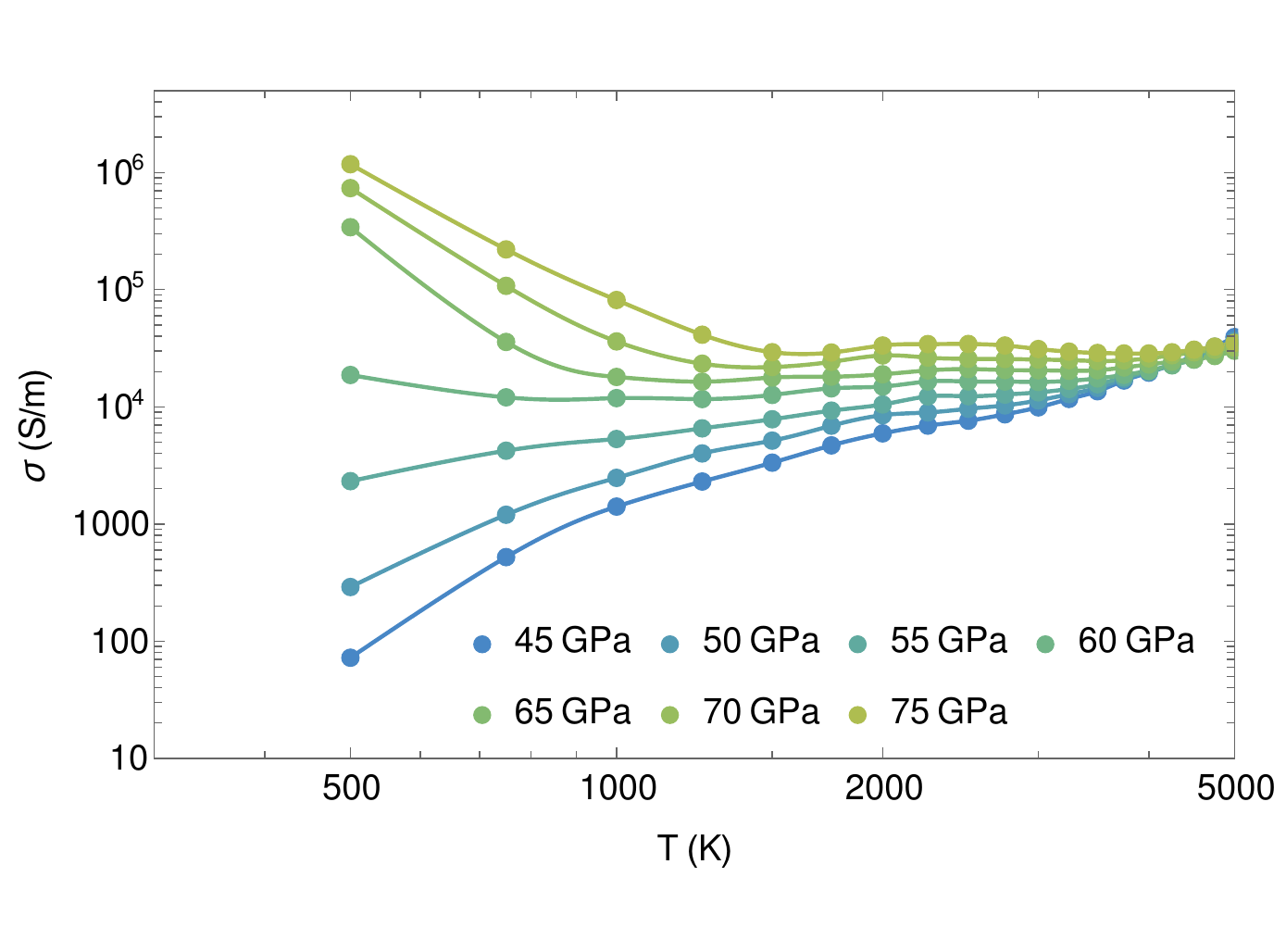}
\end{center}
\caption{Theoretical phase diagram for the $B1$ (rocksalt) phase of FeO (top panel), as a function of pressure estimated from the experimental equation of state \cite{fischer2011}. Color-coded are calculated values for electrical conductivity $\sigma$, which span only about one order of magnitude within the entire QC region at magnitudes comparable to the MIR limit ($\sim  10^5$ S/m) in other Mott oxides \cite{MIR_limit}. Solid and dashed white lines show the geotherm \cite{Wolf2015,Dobrosavljevic2022} and experimentally estimated phase boundaries \cite{fischer2011,Fischer2010}, respectively. The characteristic "fan-like" evolution of the temperature-dependent conductivity curves for 45 GPa 
$\leq  P \leq$ 75 GPa is shown in the bottom panel, as expected for Mott quantum criticality \cite{Furukawa2015,moire2021nature}. Note the markedly weak pressure and temperature dependence of the resistivity at $T > 2000$ K.}
\label{PTphasediagram}
\end{figure}

We relate our findings to existing knowledge on the experimental phase diagram of FeO by presenting our results as a function of pressure, where pressure is calculated at each volume-temperature condition using the experimentally determined equation of state for $B$1-FeO \cite{fischer2011}, as shown in Fig.~\ref{PTphasediagram}.  Here we include the experimentally estimated phase boundaries for different crystal structures, as well as the melting curve. 
We note that the phase coexistence region, where both insulating and metallic phases are present at $T < T_c \sim$ 370 K (omitted in Fig.~\ref{PTphasediagram}, see Fig.~\ref{DOSphasediagram}), is predicted to lie at the center of the experimentally estimated stability field for rhombohedrally distorted $rB$1-FeO. In addition, we observe that the Brinkman-Rice line below $\sim$ 2000 K, marking the onset of an orbitally selective metal, traces the experimentally reported $B1$-$B8$ transition boundary. These observations raise further questions regarding the relationship between insulator-metal transitions and crystal structures in strongly correlated systems, which merit further investigation but are beyond the scope of this study.

{\em Consequences for transport properties --} The three electronic phases identified for FeO in this study exhibit highly distinct transport properties (Fig.~\ref{PTphasediagram}). Conductivity in the insulating state is relatively low ($\sim 10^0 - 10^3$ S/m) and increases with temperature, as expected for thermal activation. In the correlated metallic state, conductivity is large ($\sim 10^6 - 10^8$ S/m) and decreases with increasing temperature. In contrast, conductivity in the QC state lies at intermediate levels ($\sim 10^4 - 10^5$ S/m) and displays remarkably weak dependence on both pressure and temperature. As discussed above, transport in the QC state is a consequence of a (poorly) conducting $t_{2g}$ band that lacks the presence of coherent quasiparticles. Unlike in a quasiparticle metal, where the mean-free path for electron-electron scattering is generally much longer than the lattice spacing, conductivity in the QC state lies around the Mott-Ioffe-Regel (MIR) limit ($\sim 10^5$ S/m) characterized by a short mean-free path comparable to the lattice spacing \cite{MIR_limit}. Physically, the electrons exhibit Brownian-style diffusive motion caused by strong and frequent scattering. \vspace{12pt} \\
\noindent {\em Discussion} \vspace{12pt} 
\begin{figure}[t]
\includegraphics[width=3.8in]{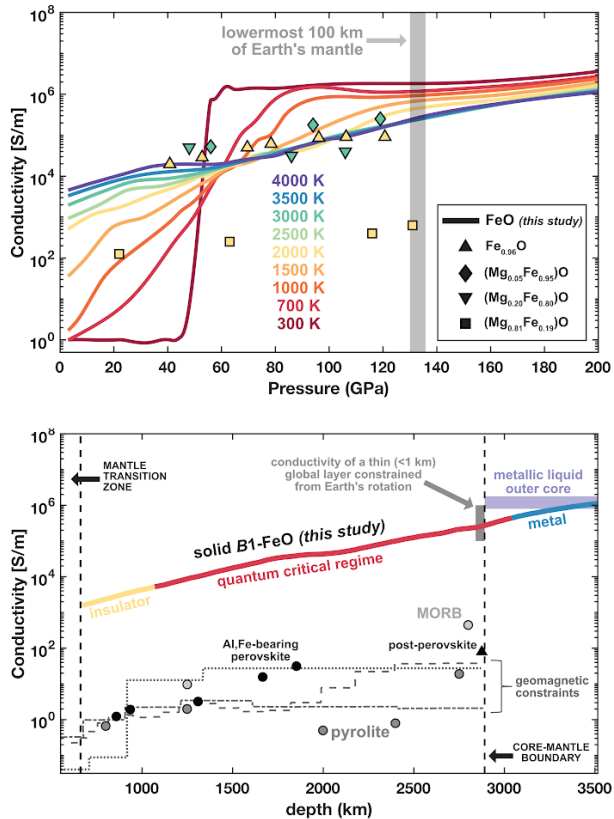}
\caption{Electrical conductivity of FeO calculated in this study (solid lines) is compared with other Earth materials as a function of pressure (top panel) and depth in the Earth along the geotherm (bottom panel). Top panel: symbols show experiments on Fe$_{0.96}$O \cite{ohta2012}, (Mg$_{0.05}$Fe$_{0.95}$)O and (Mg$_{0.20}$Fe$_{0.80}$)O \cite{Ohta2014}, and (Mg$_{0.81}$Fe$_{0.19}$)O \cite{Ohta2017}, color-coded by temperature. Bottom panel: gray lines show geomagnetic constraints on bulk mantle electrical conductivity (dashed \cite{Olsen1999}, dotted \cite{Civet2013}, dot-dash \cite{Velimsky2021}); points show experimental reports for bridgmanite (perovskite) \cite{Sinmyo2014}, post-perovskite \cite{Ohta2008}, pyrolite, and MORB \cite{Ohta2010}; blue shading shows theoretical reports for liquid iron alloys \cite{Pozzo2013,DeKoker2012,Wagle2019}; gray shading shows the conductivity range of a proposed thin layer at the mantle base \cite{Buffett2015}.}
\label{minerals}
\end{figure}

{\em Robustness of theoretical results --} The theoretical results we have obtained 
reveal that at temperatures on the scale of thousands of Kelvin, the insulator to metal crossover displays a significant intermediate regime, in close analogy to what is generally expected for quantum criticality \cite{Terletska,Vucicevic2013,Furukawa2015}. 
Although there are several aspects of our work that may shift the precise location of the crossover, the general topology of the phase diagram would not be affected. Our result was obtained for specific values of the interaction parameters $U$ and $J$, which we fixed to the values expected under ambient conditions \cite{Mandal2019}. We did so to avoid deliberate "fine-tuning" of input parameters, although we do expect that these interactions should display some volume/pressure dependence. Still, these details should not affect the qualitative and even the semi-quantitative aspects of our results. Similarly, the presence of small concentrations of Fe vacancies or small amounts of Mg substitutions could slightly displace the crossover line positions. We emphasize, however, that the characteristic scale of the electrical conductivity set by the Mott-Ioffe-Regel limit in the QC regime ($\sim 10^5$ S/m) should be a robust feature of our results. In particular, the modest pressure and temperature dependence of transport in the QC regime suggests that small shifts in the crossover line positions due to the effects discussed above will not affect the key finding that FeO exhibits intermediate values of electrical conductivity ($\sim 10^5$ S/m) at lowermost mantle conditions. Furthermore, various other physical mechanisms (such as different forms of magnetic order) that often play out at low to ambient temperatures ($T \sim 10^1 - 10^2$ K) are expected to be negligible at the $T \sim 10^3$ K levels that we consider here. In this sense, the single-site DMFT theory we adopt, which deliberately ignores such magnetic correlations, should be regarded as an accurate solution to the electronic many-body problem under conditions relevant to Earth's interior.\vspace{12pt} 

{\em Comparison to previous results --} 
We quantitatively compare general trends and magnitudes of transport obtained from our theoretical calculations to previous experimental measurements. Static compression experiments reported weak temperature dependence of electrical conductivity when heating up to $\sim 2500$ K in the pressure range $\sim$40 to 80 GPa and when heating above $\sim 2000$ K from 80 to 125 GPa \cite{ohta2012}. The QC region determined in our study spans these $P$--$T$ conditions and provides a physical basis for the observed weak temperature dependence. In addition, the same experiments reported a plateau in conductivity at around $10^5$ S/m along a pressure range of $\sim$60 to 120 GPa for $T = 1850$ K \cite{ohta2012} (Fig. \ref{minerals}). These findings strongly match the conductivities predicted in this study for the QC region, both in terms of the presence of a conductivity plateau (weak pressure dependence) and in terms of magnitudes converging around the Mott-Ioffe-Regel limit ($\sim 10^5$ S/m) (Fig. \ref{minerals}). Our findings suggest that very shallow minima in resistivity-temperature measurements from these experiments should not be interpreted as marking the location of a sharp insulator-metal transition but 
could stem from secondary effects, such as phonon (lattice) interactions or defect mobility. A shock compression study also reported conductivities on the order of $10^{5.5} - 10^6$ S/m for pressures between 72 and 155 GPa and at elevated but unconstrained temperatures \cite{Knittle1986}. Overall, the electronic phase diagram and consequent transport properties determined for $B$1-FeO in this study provide a clear physical explanation for experimental reports on the material's conductivity. Our theoretical results capture much of the same features as those reported in previous theoretical works performed by using DFT+DMFT methods for $B$1-FeO \cite{ohta2012,leonov2017prb,leonov2020prb}, although our expansive canvassing of the entire phase diagram provides qualitatively new insight and interpretation. Specifically, we demonstrated that a clearly defined intermediate regime arises between the insulator and the metal, with distinct spectral and transport signatures. 

{\em Implications for Earth's interior --} We find that the electrical conductivity of FeO at lower mantle conditions exhibits intermediate values ($10^4-10^5$ S/m) relative to the insulating mantle and metallic core. The lower mantle features conductivity magnitudes of $10^0$ to $10^2$ S/m based on experimental and geomagnetic constraints. Experiments on the major lower mantle phases bridgmanite \cite{Sinmyo2014}, ferropericlase \cite{Ohta2017}, and post-perovskite \cite{Ohta2008} have reported conductivities from $10^0$ to at most $10^3$ S/m, similar to experiments on the hydrous silicate phase D \cite{Li1991}, as well as on pyrolite and mid-ocean ridge basalt (MORB) rocks \cite{Ohta2010} that represent the average lower mantle and subducted oceanic crust, respectively (Fig. \ref{minerals}). These values are in good agreement with depth profiles for conductivity, determined from geomagnetic observations \cite{Velimsky2021,Olsen1999,Civet2013}. For the metallic outer core, theoretical computations have reported conductivities for liquid iron alloys around $10^6$ S/m \cite{Wagle2019,DeKoker2012,Pozzo2013}, similar to experimental measurements on solid iron and iron alloys at high $P$--$T$ conditions \cite{Zhang2020,Zhang2021,Inoue2020,Ohta2016}. The intermediate conductivity values for FeO at lowermost mantle conditions ($\sim10^5$ S/m) are robust even for small amounts of Mg substitution (up to 20\%), based on experimental results \cite{Ohta2014}, suggesting that iron-rich (Mg,Fe)O in the lowermost mantle would exhibit a unique signature of electrical conductivity relative to coexisting materials. 

The presence of solid FeO-rich regions has recently been quantitatively supported from combined seismological, geodynamic, and mineralogical constraints showing that ultralow velocity zones at the core-mantle boundary can be explained by the presence of highly iron-rich (Mg,Fe)O \cite{Lai2022,Dobrosavljevic2019,Wicks2017}. These structures have been most abundantly discovered beneath mantle plumes and around LLSVPs beneath the Pacific and Africa \cite{Yu2018,Kim2020,Jenkins2021,Cottaar2022}. Geodynamic work has further suggested that these mountain-scale structures may form from a thin layer \cite{Bower2011,Li2017a} that could be difficult to detect seismically \cite{Russell2022}. The bulk conductivity of such features would depend on the interconnection of moderately conductive FeO in the assemblage, which is poorly constrained. However, the remarkably low viscosity of the material ($10^{12}$ Pa-s) at lowermost mantle conditions \cite{Reali2019} and its relatively high abundance in ultralow velocity zones suggested by recent work ($\sim 20-40$\%) \cite{Lai2022,Dobrosavljevic2019,Jackson2021} supports the possibility of interconnected networks of iron-rich (Mg,Fe)O and resulting bulk conductivity similar to $10^5$ S/m. 

This finding is particularly intriguing due to several independent lines of observational evidence pointing to a thin layer of moderately conductive material at the mantle base that affects the electromagnetic coupling of the core and mantle and thus Earth's rotation and magnetic field. Specifically, variations in the length of day over periods of several decades, as well as nutations of Earth's rotation axis on the diurnal timescale, are best explained by a mantle basal layer ~1 km thick with conductivity $10^5$ S/m \cite{Buffett2015}. Further, low temporal variations of Earth's magnetic field in the Pacific region have recently been attributed to a non-uniform conducting layer at the mantle base with higher conductance levels in the Pacific, estimated at $6-9\times 10^8$ S compared to $10^8$ S for a global average layer \cite{Dumberry2020}. This elevated conductance could be approximately explained by 20-30 km thick structures with conductivity $\sim10^5$ S/m covering around one-third of the mantle base on the Pacific, compatible with typical heights and detection locations of ultralow velocity zones \cite{Yu2018}. Our theoretical results and previous experiments show that FeO enrichment could provide a strong explanation for the inferred thin moderately conductive layer and/or moderately conductive structures at Earth's mantle base.

A solid FeO-rich mineralogy could thus provide a unifying explanation for both seismic observations of ultralow velocity zones as well as independent observations of temporal variations in Earth's rotation and magnetic field. FeO-rich structures could further imply heterogeneous thermal conductivity at the core-mantle boundary, instead of homogeneous heat flow out of the core assumed in some models of mantle dynamics \cite{Buffett2007,Lay2008}. Using the Wiedemann-Franz law and our calculated conductivity of $\sim 2\times 10^5$ S/m, we estimate an electrical contribution to the thermal conductivity of $\sim$17 W/m-K for FeO at the core-mantle boundary. This value is around two to four times larger than the reported thermal conductivity of the average pyrolitic lowermost mantle \cite{Geballe2020}. Solid FeO-rich ultralow velocity zones may thus represent regions of high thermal conductivity at Earth's mantle base, which could promote the generation of long-lived mantle plumes, influence convection dynamics, and affect crystallization processes in the core
\cite{Olson2010,Driscoll2011,Wang2022}. \vspace{12pt}

 {\em Methods --} 
The eDMFT algorithm we use \cite{Georges1996,dmft2006rmp,haule2010} starts with the calculation of the eigen-energies and the eigen-wavefunctions of the crystal by solving the DFT equations. Next, the correlated orbital subset is projected out as "quantum impurities" by a real-space projectors without downfolding, while the uncorrelated orbitals are treated by DFT, and act as   a mean-field bath on the quantum impurities, resulting in a hybridization between the two.  The hybridization functions are determined self-consistently by solving the DMFT equation. The quantum impurities are solved by the hybridization expansion continuous time quantum Monte Carlo (CTQMC) \cite{Werner2006,Haule2007} method.  The modified charge density derived from combined DFT and DMFT equations is then used as the input of the next DFT iteration. The eDMFT algorithm iterates until full convergence of the charge density is achieved, the impurity self-energies, and the lattice Green's function. Finally, the maximum entropy method \cite{MEM} is employed to analytically continue the Green's function and the self-energy from the Matsubara frequency to the real frequency axis. The linear augmented plane wave method is used as a basis,  as implemented in WIEN2K package \cite{Blaha2001}, and the local density approximation \cite{LDA}  to the exchange and correlation functional is employed in the DFT part. The correlations treated by both the DFT and eDMFT are subtracted exactly \cite{exactDC}.  In each DMFT iteration a huge number ($\sim$ 2.8$\times10^{10}$) of Monte Carlo updates is used to reduce the statistical error. A Monkhorst-Pack mesh of at least $12\times12\times12$   $k-$ points is used in the calculation.  At the ambient pressure the energy window for projection of the correlated states is $\pm 10$ eV around the Fermi energy. At high pressure the energy window is expanded so that the same number of bands are included for projection as done at ambient pressure. Only the Fe-$3d$ electrons are treated as correlated with Coulomb interaction $U$ = 10.0 eV and Hund's coupling $J$ = 1.0 eV, which is based on previous constrained DMFT calculations of FeO at ambient pressure \cite{Mandal2019}. Throughout the paper we fix the Coulomb interaction $U$ and the Hund's coupling $J$ as volume independent. Although increased pressure should reduce $U$ and $J$ in real FeO material, it will only quantitatively tune the results in the paper, such as the exact position of the insulator-metal transition.  

{\em Author contributions --}  V.D. designed the  project. V.V.D. and J.M.J. provided the geophysical context and implications. W.D.H. and P.Z. (co-first authors) equally contributed to the computational work and data analysis. K.H. provided the eDMFT  code and technical guidance. W.D.H. and V.V.D. produced the figures. V.V.D. and V.D. wrote the original manuscript. All authors discussed the results and commented on the manuscript.

 \noindent {\em Acknowledgements -- }  Work in Florida (W.D.H. and V.D.) was supported by the NSF Grant No. DMR-1822258, and the National High Magnetic Field Laboratory through the NSF Cooperative Agreement No. DMR-1644779 and the State of Florida. Work at Rutgers (K.H.) was supported by the NSF grant No. DMR-2233892. P.Z. acknowledges the support of NSFC grant No.11604255. J.M.J. and V.V.D. are grateful to the National Science Foundation's Collaborative Study of Earth's Deep Interior (EAR-2009935) and Geophysics (EAR-1727020) programs for support of this work.
 
 {\em Data Availability -- } The theoretical data presented in the figures can be found in the Source Data files, which are provided with this paper. The full set of theoretical data generated during this study are available from the corresponding author upon reasonable request.

\bibliographystyle{apsrev}

\newpage

\section*{Supplementary Materials:}

\section{Thermal destruction of quasiparticles and the Brinkman-Rice line}

According to Fermi liquid theory \cite{pinesnozieres}, even strongly correlated metals should qualitatively behave similarly to an ideal gas of fermions, but only concerning sufficiently low-energy excitations. Here thermodynamic response, as well as transport behavior, is dominated by low-energy {\em quasiparticle (QP) excitations}. In the presence of strong electronic correlations, these QPs still carry charge {\em e} and spin 1/2, but often feature a significantly enhanced effective mass $m^*$. The corresponding electronic states assume a form of a narrow {\em QP band}, which produces a sharp {\em QP peak} \cite{Georges1996} in the density of states (DOS), around the Fermi energy. The small spectral weight $Z \sim 1/m^*$ of these QP peaks encodes the extreme fragility of such correlated matter to thermal excitations. 

Unlike conventional metals, with characteristic energy on the scale of the Fermi temperature $T_F \sim 10^4$ K, the QP states found in strongly correlated materials can be dramatically affected by much lower temperatures, modifying all observable properties. The characteristic energy scale of these QP states was first estimated theoretically by Brinkman and Rice \cite{brinkmann70prb}, based on the Gutzwiller variational approch. The corresponding temperature $T_{BR}$, at which the QP states are thermally destroyed, is often called the "Brinkman-Rice temperature" \cite{rqp2013prl}. It marks the crossover from the coherent regime dominated by long-lived QP excitations to an incoherent regime, dominated by very strong electron-electron scattering, which we identify with a "quantum critical" (QC) regime \cite{Terletska,Furukawa2015,moire2021nature} associated with the Mott metal-insulator transition. 

\begin{figure}[b]
\includegraphics[width=3in]{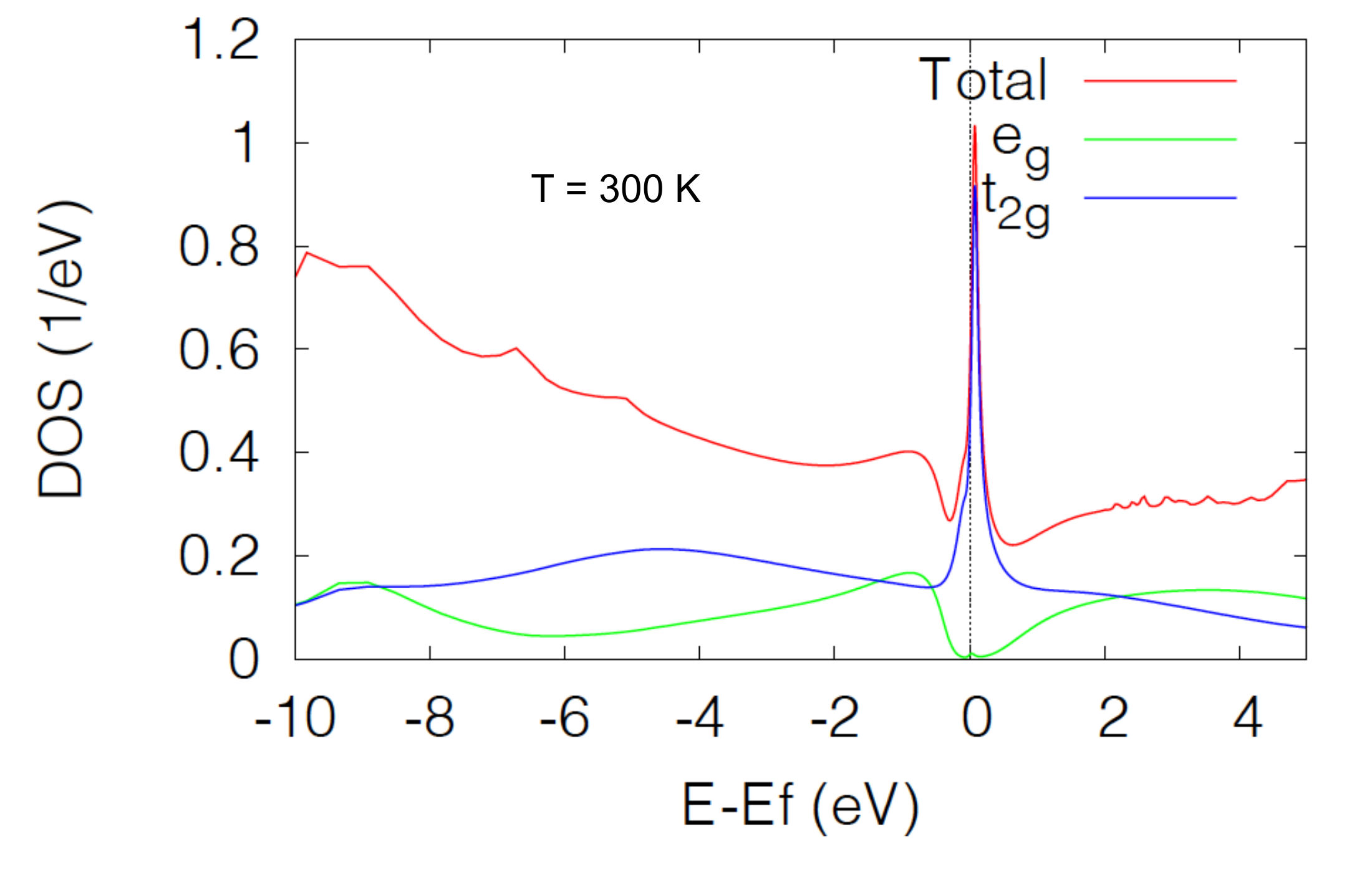}
\includegraphics[width=3in]{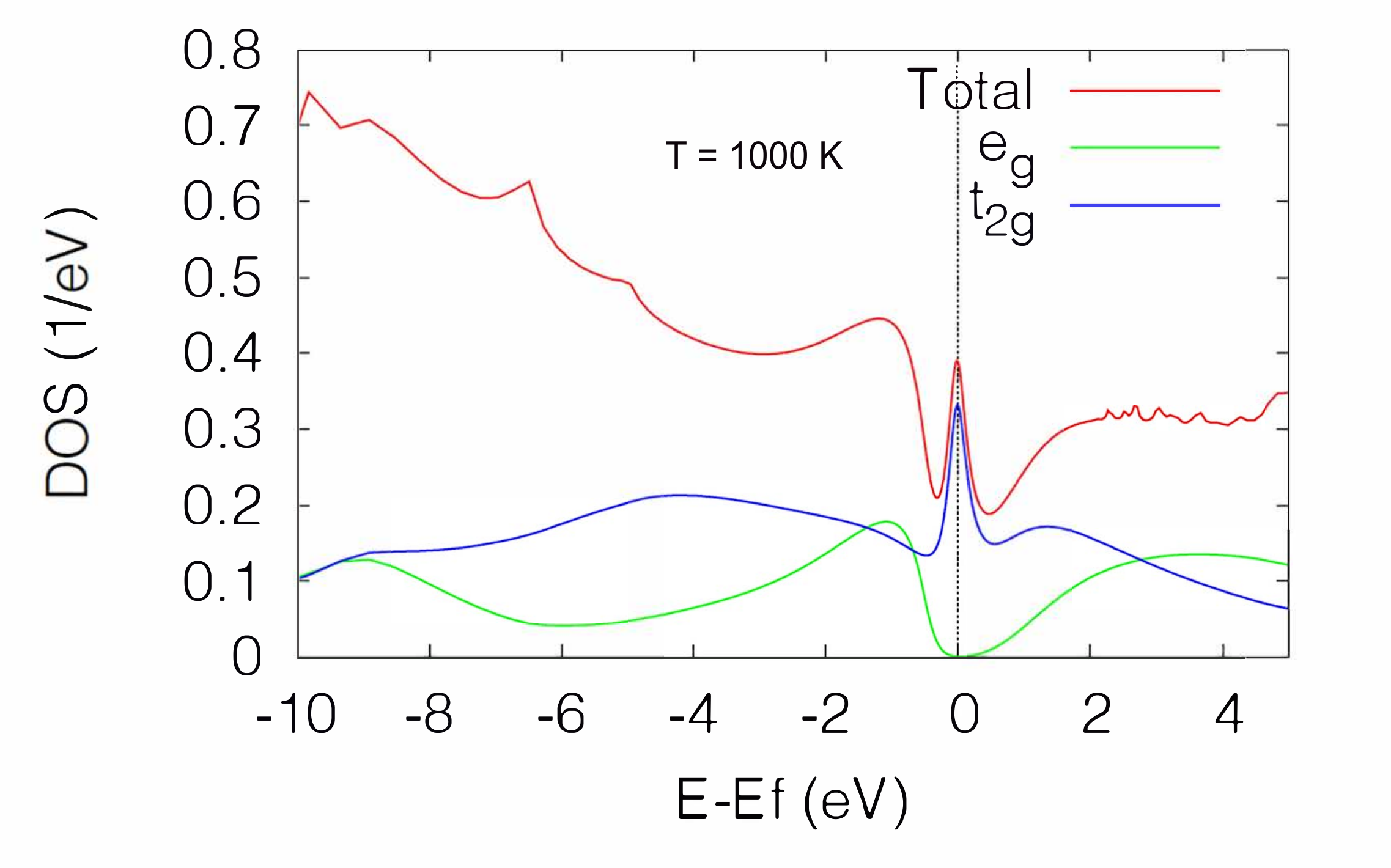}
\includegraphics[width=3in]{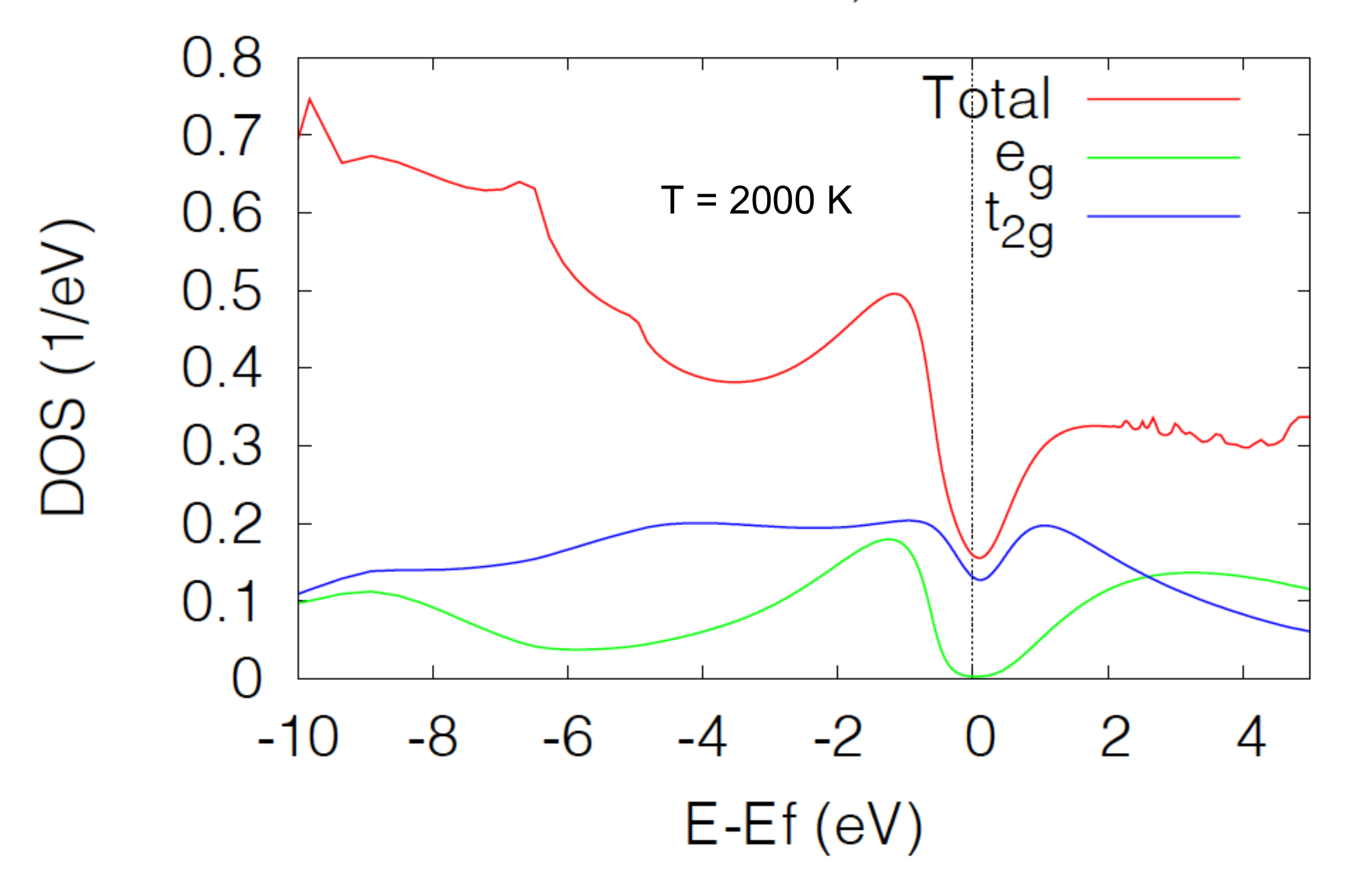}
\caption{Electronic DOS of FeO on barely the metallic side of the Mott point ($\Delta \rm{v} = 0.25$). The top panel shows the spectrum at $T$ = 300K, featuring a sharp QP peak at the Fermi energy, with modest spectral weight. Increasing temperature quickly destabilizes/broadens the fragile QP states, as shown for $T$ = 1000 K (middle panel). The QP states are completely suppressed above the Brinkman-Rice temperature (which is $T_{BR} \approx 1100 K$ for this compression), marking the crossover to the QC region at higher temperatures. Here the $t_{2g}$ electronic states are incoherent but already gapless, while the $e_g$ gap remains open, as shown for $T$ = 2000K (bottom panel).}
 \label{BR}
\end{figure}

Most existing theoretical approaches \cite{brinkmann70prb} to correlated electronic matter focus on describing the ground state and only the leading low-temperature excitations. They therefore cannot properly describe this coherence-incoherence crossover around $ T \sim T_{BR}$, and the physical properties in the regime where thermal excitations dramatically affect the electronic spectra. In contrast, the new theoretical methods based on DMFT and its extensions \cite{Georges1996,dmft2006rmp} are hand-tailored precisely to self-consistently determine the key player in this regime -- the electron-electron scattering rate. Physically, when this scattering rate becomes comparable to the characteristic energy scale of quasiparticles, the corresponding electronic states are thermally suppressed. The sharp quasiparticle peak in the DOS spectra then "melts away" \cite{MIR_limit,rqp2013prl} in a fairly sudden fashion, which happens at $ T \sim T_{BR}$. This behavior can be clearly seen in Fig.~\ref{BR}, where we show the evolution of DOS spectra with increasing temperature, focusing on the "orbitally selective" metal \cite{hund2013review} which forms at low temperature as soon as the $t_{2g}$ gap closes. While a very sharp QP peak is seen at $T = 300$ K, it is quickly diminished already at $T = 1000$ K, and it completely disappears at $T= 2000$ K, where it is replaced by a shallow pseudogap feature around the Fermi energy (for comparison with similar behavior in a one-band Hubbard model, see Fig.~7 in Ref.~\cite{Vucicevic2013}).  

\begin{figure}[t]
\includegraphics[width=3in]{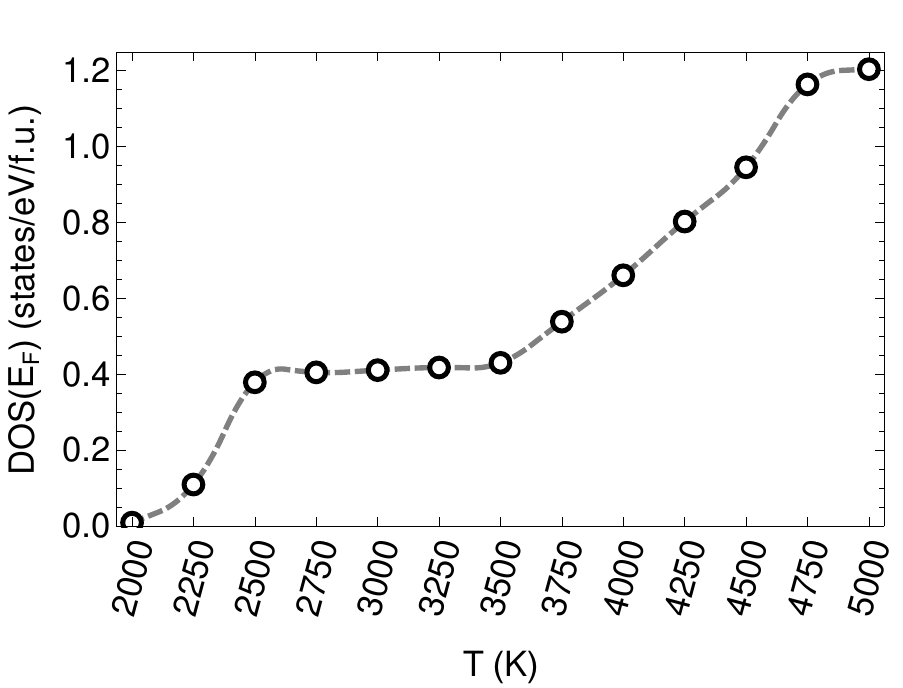}
\includegraphics[width=3.15in]{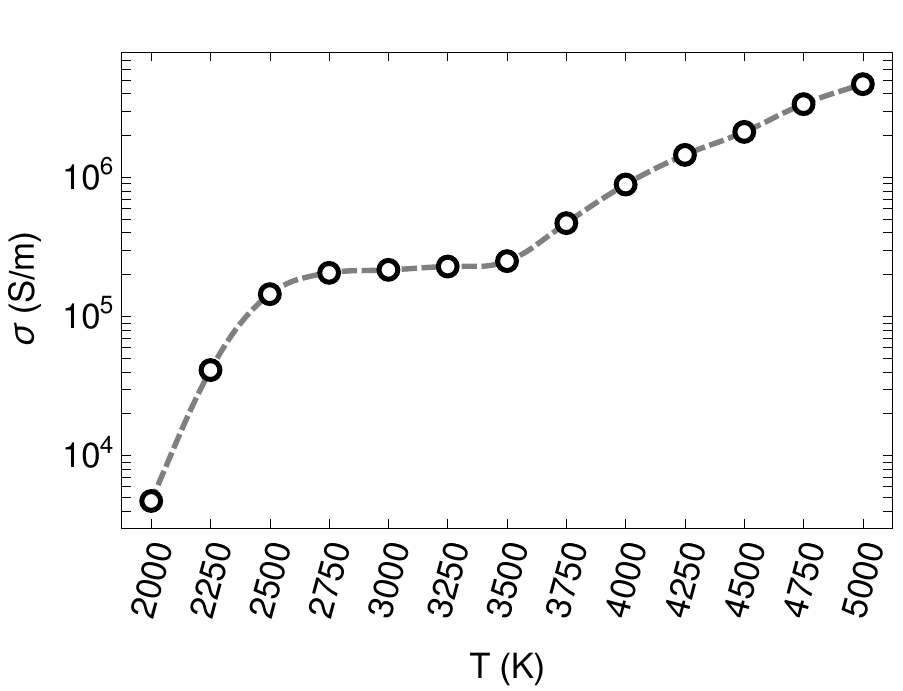}
\caption{Both DOS (left panel) and the electrical conductivity (right panel) plotted along the geotherm trajectory. Both display remarkably weak $T$-dependence within the lower mantle, just outside the core-mantle boundary (2500 K < $T$ < 3500 K).}
 \label{geotherm}
\end{figure}

The sudden modification of spectral features with temperature allows us to identify the corresponding crossover temperature $T_{BR}$, which we define as the temperature where the DOS value at the Fermi energy (the "height" of the QP peak) is suppressed. This procedure has been repeated at each of the values of compression studied, defining the "BR line" we displayed on Fig.~1 (main text). We should emphasize that this BR line is {\em not} a sharp phase transition but is only a (relatively smooth) crossover between two physically distinct regimes. In our case, it marks the boundary of the incoherent QC regime, which differs significantly from the low-temperature quasiparticle metal. Note that $T_{BR}$ increases with compression (see Fig.~1 of main text), so that the destruction of QP states can also be observed by reducing compression at fixed temperature in such a way to cross the BR line (see Fig.~2 of main text, bottom panels). This strategy is especially useful at higher temperatures ($T$ > 2000 K), where the BR line is seen to suddenly "turn up", due to the enhanced metallization caused by the closing of the $e_g$ gap around $\Delta \rm{v} = 0.34$. It is interesting to observe how the thermal evolution of both the electronic spectra (see color-coded DOS in Fig.~1 of main text) and the electrical conductivity (see color-coded conductivity in Fig.~4 of main text) closely tracks the BR line in the entire phase diagram. The dramatic "upturn" of the BR line at $T > 2000$ K makes it almost "vertical" on the phase diagram. The net result of all this is a notably weak temperature dependence of all quantities within the QC phase. Remarkably, the geotherm trajectory also is almost "vertical" (i.e. pressure-independent) in the lower mantle region just outside the core-mantle boundary. As a result, both the DOS and the electrical conductivity display almost no $T$-dependence when plotted along the geotherm line (at 2500 K < $T$ < 3500 K), as shown in Fig.~\ref{geotherm}. 

We should also mention the often-discussed "Fermi liquid regime" (FL), associated with the {\em leading} low-temperature behavior of quasiparticles, where the conductivity $\sigma (T) \sim T^{-2}$. It is worth stressing that this regime corresponds to $T <T_{FL}$, with $T_{FL}$ that is usually significantly smaller than $T_{BR}$. The intermediate regime $T_{FL}  < T < T_{BR}$ is sometimes described as featuring {\em "resilient QPs"} \cite{rqp2013prl}, where the QP parameters (e.g. the QP weight $Z$) assume a certain $T$-dependence, and other properties display deviations from standard FL behavior. Both temperature scales $T_{FL}$ and $T_{BR}$ have been experimentally \cite{pustogow2021npjQM} and theoretically \cite{rqp2013prl,Vucicevic2013} identified in certain Mott systems, and are both seen to decrease towards the Mott transition. In this study, however, we shall not explore the details of such low-$T$ QP behavior, primarily because the corresponding correlated metallic $B1$-FeO phase is not of much direct relevance to the issues surrounding the core-mantle boundary.

\section{Estimating the Mott gap}

At ambient conditions FeO is a robust Mott insulator, with a substantial gap to electronic excitations on the scale of several eV. In systems with cubic symmetry (corresponding to the $B1$ {\em rocksalt} structure), the $d$-electrons of Fe are distributed between $t_{2g}$ and $e_g$ orbitals which, within a solid, contribute to forming the "Hubbard" bands with corresponding symmetry. Because crystal field splitting lifts the degeneracy between these orbitals/bands, the corresponding Mott gaps are also in-equivalent. In FeO the $t_{2g}$ band gap is generally smaller than the $e_g$ gap, and it closes first under compression around  $\Delta \rm{v} \approx 0.22$; the $e_g$ gap closes around $\Delta \rm{v} \approx 0.43$.

\begin{figure}[h]
\includegraphics[width=3in]{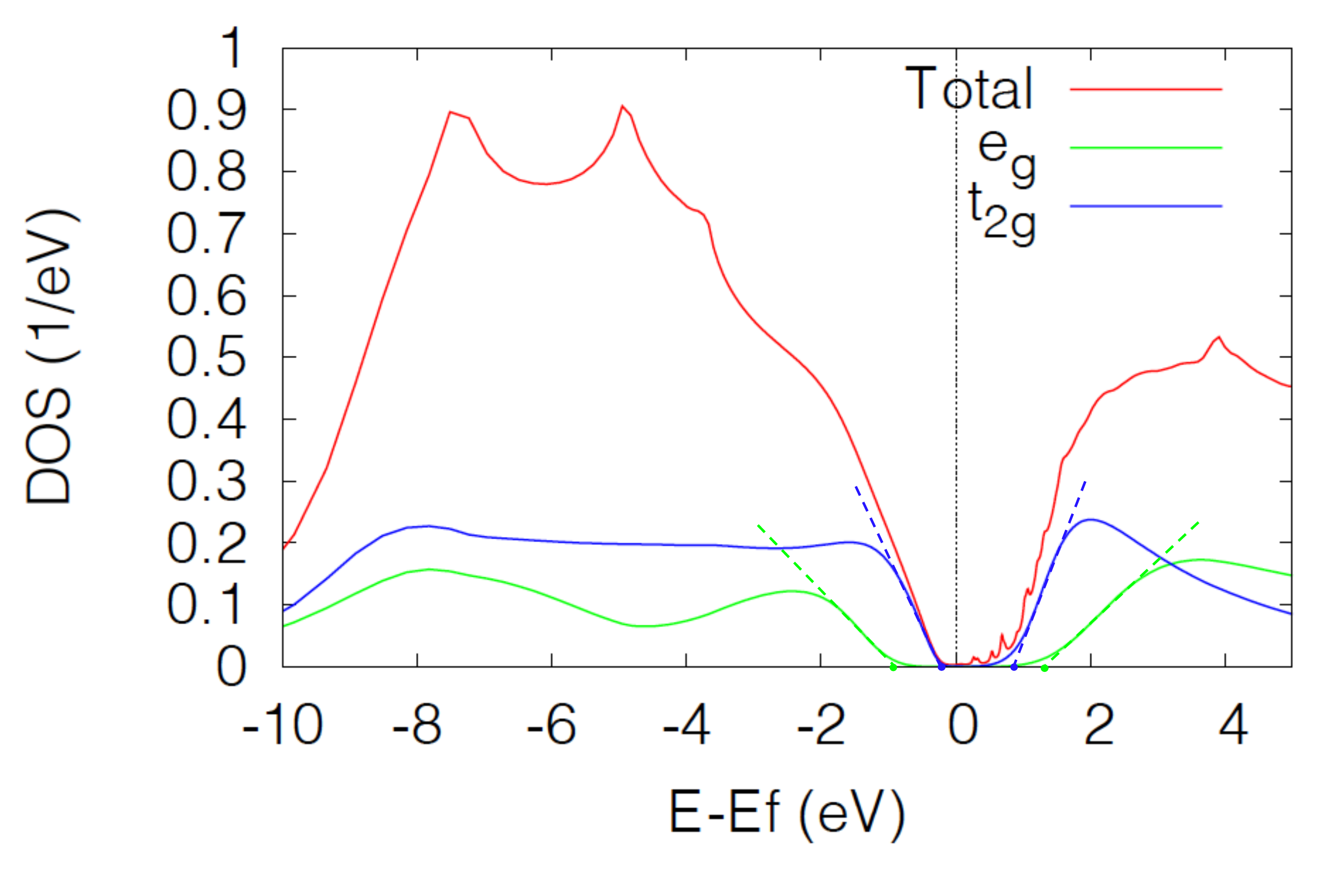}
\caption{Density of states (DOS) for FeO under ambient conditions ($P=0$ GPa, $T=300$ K) featuring substantial gaps both in the $t_{2g}$ and the $e_g$ orbital/band. The estimate for the low-temperature gap size is performed by linearly extrapolating DOS to zero, towards the band edge.}
 \label{gap}
\end{figure}

In order to understand the approach to the IMT, as well as its finite-temperature manifestations, it is important to precisely determine the evolution of these band gaps with compression. The precise definition of the band gap is, however, a bit complicated in our finite-temperature calculations, since thermal excitations generally tend to create band tails and "gap rounding". Still, when the gap size $E_g$ is large as compared to the thermal energy $k_B T$, this effect is modest and a relatively accurate estimation of the $T=0$ value of the gap is possible. To obtain a quantitative estimate despite such thermal "rounding", we adopt the procedure to linearly extrapolate DOS of a given band, a procedure that can roughly eliminate the rounding effect. This procedure is justified by the fact that, within our DMFT-type setup, the DOS has sharp band edges at $T=0$; it has also been cross-checked for simple model systems \cite{Vucicevic2013}, where in certain limits the gap size is accurately know. 

To illustrate this procedure, in Fig.~\ref{gap} we show how the extrapolation is performed under ambient conditions. This procedure has been repeated for all the values of compression considered, using the results obtained at $T=300$K, and we obtain the low-temperature value of the given gap energy $E_g$, as a function of compression. On physical ground and based on previous extensive work on model systems \cite{Terletska,Vucicevic2013}, we expect that typical  insulating behavior should be observed only at temperatures substantially smaller than the thermal energy corresponding to the gap size. Thus, we define $T_{gap} = E_g /k_B$ as an appropriate temperature scale that marks the boundary of the insulating region, separating it from the intermediate {\em gapless} QC region, which displays incoherent transport at finite temperature. Since $E_g$ decreases linearly with compression (for both band gaps), so does the corresponding crossover scale $T_{gap}$, which vanishes where the $T=0$ gap does. 

This expectation is directly confirmed by plotting $T_{gap}$ as a function of compression on Fig.~1 (main text). Here the DOS at the Fermi energy obtained by our expansive direct computations across the phase diagram is color coded; we observe how DOS assumes (exponentially) small values in the entire region where $T < T_{gap} (\Delta\rm{v})$, for given compression. Similar results are obtained in Fig.~4 (main text), where the electrical conductivity is color-coded across the phase diagram, as a function of pressure and temperature. Again, we observe how the conductivity is (exponentially) small in the Mott insulating phase, corresponding to $T < T_{gap} (P)$.

\section{Data grid}

\begin{figure}[h]
\includegraphics[width=3.3in]{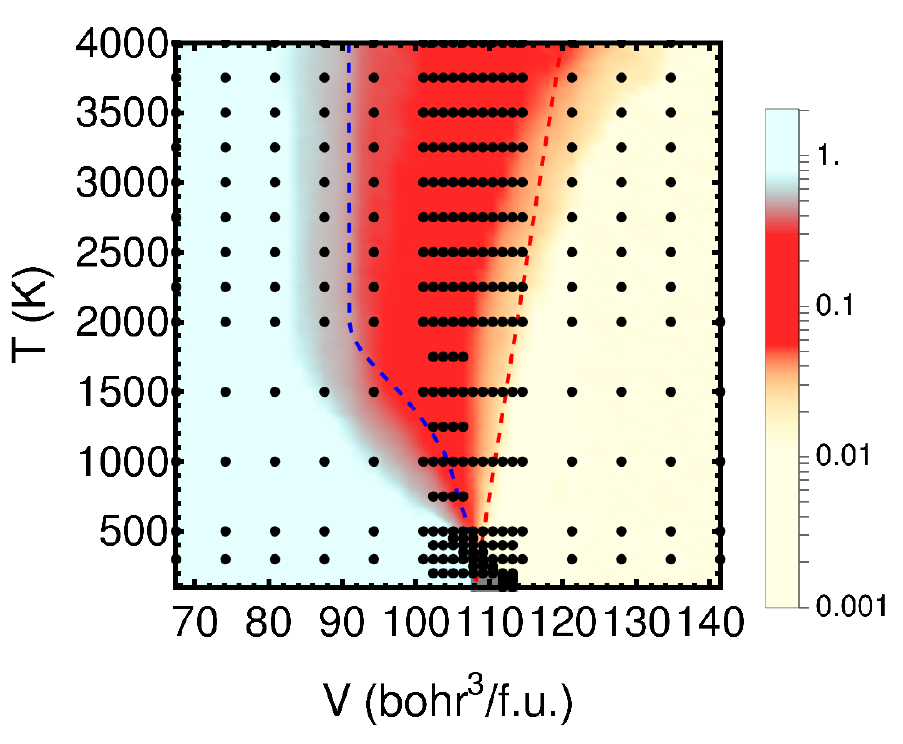}
\caption{Phase diagram of $B1$-FeO, with color-coded values of DOS at the Fermi energy. Black dots indicate the grid of points where we solved the eDMFT equations.}
 \label{grid}
\end{figure}

While most studies in the past obtained accurate DFT+DMFT data only for a few values of temperatures and volume (pressure), we have performed an expansive set of calculations, canvassing the entire phase diagram of $B1$-FeO. We solved the eDMFT equations on a dense grid of points across the insulator to metal transition (IMT) region. In doing so, we selected a finer grid around the critical volume, since this is where all physical quantities display a somewhat stronger dependence on volume/pressure, as shown on Fig.~\ref{grid}. The color coded values of DOS and the electrical conductivity were obtained by appropriate spline procedures to interpolate between the results explicitly obtained at these data points.

\bibliographystyle{apsrev}

\end{document}